\documentclass{webofc}
\usepackage[varg]{txfonts}   
\begin{document}
\title{Understanding Shape and Centroid Deviations in 39 Strong Lensing Galaxy Clusters in Various Dynamical States}
%
%

\author{\lastname{Raven Gassis}\inst{1}\fnsep\thanks{gassismr@mail.uc.edu} \and
        \lastname{Matthew B. Bayliss}\inst{1} \and
        \lastname{Keren Sharon}\inst{2} \and
        \lastname{Guillaume Mahler}\inst{3} \and
        \lastname{Michael D. Gladders}\inst{4} \and
        \lastname{Håkon Dahle}\inst{5} \and
        \lastname{Michael K. Florian}\inst{6} \and
        \lastname{Jane R. Rigby}\inst{7} \and
        \lastname{Michael McDonald}\inst{8} \and
        \lastname{Lauren Elicker}\inst{1} \and
        \lastname{M. Riley Owens}\inst{1} 
}

\institute{Department of Physics, University of Cincinnati 
\and
           Department of Astronomy, University of Michigan 
\and
           Centre for Extragalactic Astronomy, Durham University
\and
            Department of Astronomy and Astrophysics/Kavli Institute for Cosmological Physics, University of Chicago
\and
           Institute of Theoretical Astrophysics, University of Oslo 
\and
           Steward Observatory 
\and
           University of Arizona, Observational Cosmology Lab, Code 665, NASA Goddard Space Flight Center
\and
           Department of Physics/Kavli Institute for Astrophysics and Space Research, Massachusetts Institute of Technology
          }

\abstract{%
Through observational tests of strong lensing galaxy clusters, we can test simulation derived structure predictions that follow from $\Lambda$ Cold Dark Matter ($\Lambda$CDM) cosmology. The shape and centroid deviations between the total matter distribution, stellar matter distributions, and hot intracluster gas distribution serve as an observational test of these theoretical structure predictions. We measure the position angles, ellipticities, and locations/centroids of the brightest cluster galaxy (BCG), intracluster light (ICL), the hot intracluster medium (ICM), and the core lensing mass for a sample of strong lensing galaxy clusters from the SDSS Giant Arcs Survey (SGAS). We utilize \emph{HST} WFC3/IR imaging data to measure the shapes/centroids of the ICL and BCG distributions and use \emph{Chandra} ACIS-I X-ray data to measure the shapes/centroids of ICM. Additionally, we measure the concentration parameter (\emph{c}) and asymmetry parameter (\emph{A}) to incorporate cluster dynamical state into our analysis. Using this multicomponent approach, we attempt to constrain the astrophysics of our strong lensing cluster sample and evaluate the different components in terms of their ability to trace out the DM halo of clusters in various dynamical states.
}

\maketitle
\section{Introduction}
\label{intro}

Galaxy clusters form at the nodes of our Universe's cosmic web. As detailed in $\Lambda$ Cold Dark Matter ($\Lambda$CDM) physics, these galaxy clusters form via the gradual accretion and incorporation of originally separate halo systems \cite{Beers1983}. Through this process of hierarchical mergers, they are the most massive self-gravitating objects in the known universe \cite{Djorgovski1987}.

In an idealized system where the cluster is unaffected by interfering astrophysical phenomena at all scales, the brightest cluster galaxy (BCG), intracluster light (ICL), and the hot intracluster medium (ICM) should align with the DM halo of a galaxy cluster defined by its core lensing mass \cite{Sastry1968,Binggeli1982,Bosch2005,Hashimoto2008,Niederste-Ostholt2010,Biernacka2015,Donahue2016,Wang2018}. 

Even when we incorporate the complex astrophysical landscape of galaxy clusters, we still expect the differences between these distributions to be small and infrequent. However, previous work has found that some of the mass components are not always aligned with the DM halo or with each other \cite{Bosch2005,Sanderson2009,Skibba2011,Zitrin2012,Hikage2013,Lauer2014,Oliva-Altamirano2014,Wang2014,Hoshino2015,Rossetti2016,Lange2018,Lopes2018,Zenteno2020,Propis2021,Martel2014}. Utilizing the constraining power of strong lensing mass models, we can more accurately measure the centroid, shape, and orientation of the DM dominated gravitational potential to more accurately quantify the frequency of these deviations. In some cases, this is due to the disturbed nature of the cluster \cite{Propis2021}, though the frequency for which deviations occur in relaxed vs. disturbed clusters is largely unexplored in the current literature.

Given enough time to relax the BCG and ICL should revert to their shared orientation and centroid with their DM halo \cite{Montes2018,Wittman2019}. Still, Kim et al. 2017 \cite{Kim2017} and Harvey et al. 2017 \cite{Harvey2017} found that relaxed clusters can have BCGs that show residual "wobbling." This is beyond $\Lambda$CDM predictions. Though the ICL has not been found to "wobble" in the same way, the effect of dynamical disruptions on the ICL’s ability to trace out the shape of the DM halo can be tested using DM halo models derived from strong lensing. 

The ICM gas can show extreme misalignments for disturbed clusters. Additionally, hydrodynamical gas oscillations in relaxed clusters can cause some deviations in the ICM to persist even when the stellar components have realigned with the core lensing mass \cite{Markevitch2001,Churazov2003,Johnson2012,Harvey2017}.

\begin{figure*}
\begin{center}
    
	\includegraphics[width=\textwidth]{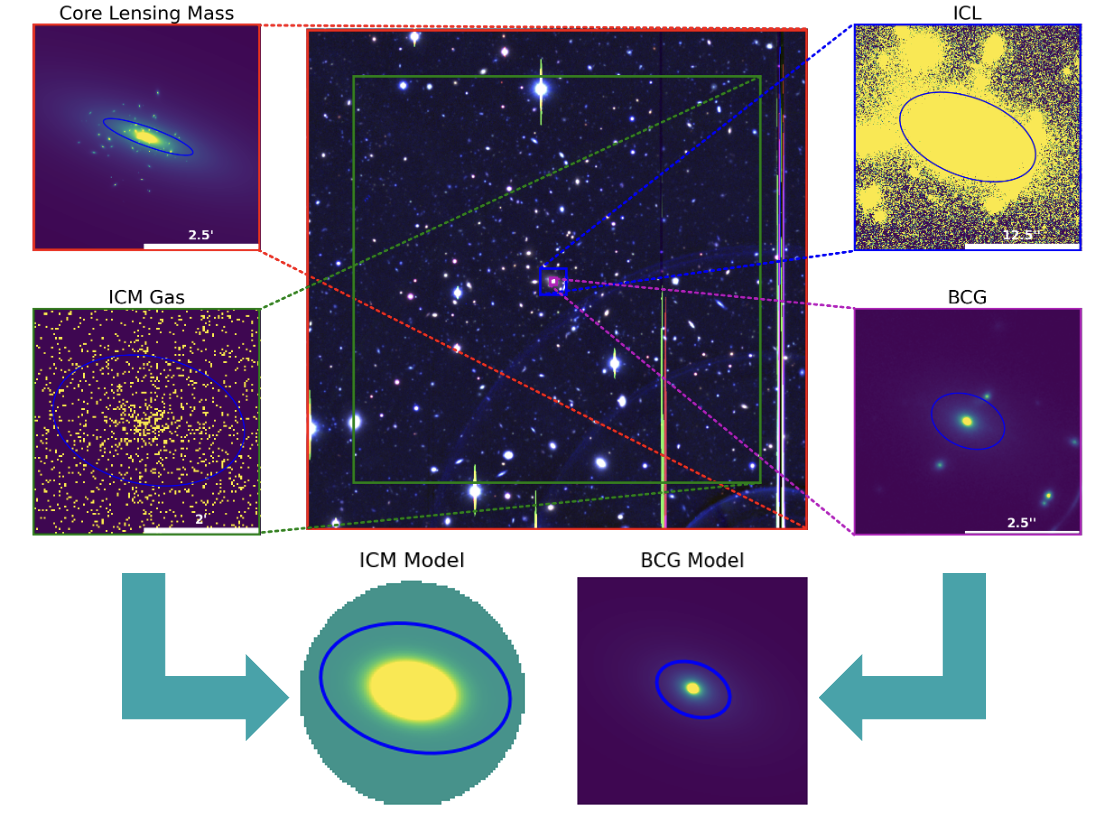}
    \caption{Visualization of the data distributions with the corresponding best fit ellipse for example galaxy cluster J0957p0509. We include a 5"x5" representative color (gri) Subaru image of the field in the center panel for a scale comparison of the different distributions. The ICM and BCG also have visualizations of their models that precede ellipse fitting.}
    \label{fig:data}
\end{center}
\end{figure*}

\section{Data and Measurements}
\label{sec:datnmeas}

In this work, we measure the shape and centroid of the BCG, ICL, and core lensing mass derived from strong lensing for 39 clusters from the Sloan Giant Arcs Survey (SGAS). Each sample cluster has a corresponding well defined lens model derived from multiband \emph{HST} data and spectroscopic data \cite{Sharon2020,Sharon2022a,Sharon2022b}. A subset of 27 clusters have \emph{Chandra} ACIS-I X-ray data that allow us to make measurements of the ICM in tandem with the other 3 components.


\subsection{BCG/ICL Measurement}

In order to isolate the BCG, we modeled the core using \texttt{GALFIT}. To separate the ICL from all other objects, we used \texttt{SExtractor} to derive masks for all objects in the field. For both distributions, we use iterative elliptical isophote fitting applying the Python \texttt{lsq-ellipse} function.



\subsection{ICM Measurement}
\label{sec:icm}

We start by using the publicly available \texttt{CIAO} tools to process our \emph{Chandra} X-ray event data. The data was point-source subtracted, binned by 2”, and set to only include events in the broad energy band (0.5-7 keV with an effective energy of 2.3 keV). The resulting processed ICM distribution was modeled using the \texttt{CIAO Sherpa} modeling functions, applying 2 elliptical Gaussian distributions to model the ICM while simultaneously fitting the background. The \texttt{cash} fitting statistic \cite{Cash1979} is used to derive the best fit parameters and 1 sigma deviation for each object.

\subsection{Dynamical State Measurement}
\label{sec:ds-meas}

We measure both the concentration and asymmetry of each of our clusters in order to understand their dynamical states. Both these parameters utilize the reduced \emph{Chandra} data described in~\ref{sec:icm}.

We define concentration by equation~\ref{eq:c}.

\begin{equation}c_{[\mathrm{R500}]}=\frac{Flux(\mathrm{r}<0.2*\mathrm{R500})}{Flux(\mathrm{r}<\mathrm{R500})}.
	\label{eq:c}
\end{equation}

Any object with a concentration measurement $c > 0.25$ is considered relaxed whereas any object with a concentration measurement $c < 0.3$ is disturbed \citep{Rasia2013}. Objects that fall between $0.25 < c < 0.3$ require additional relaxation proxy measurements.

In addition to concentration, we measure the asymmetry found by rotating the initial X-ray distribution $I$ by $180^\circ$ and subtracting the rotated distribution $R$ from the original distribution as illustrated by equation~\ref{eq:A} \citep{Rasia2013}.

\begin{equation}
A_{180}=\frac{\sum{(|I-R|)}}{\sum{I}}
\label{eq:A}
\end{equation}

Objects with an asymmetry parameter $A_{180} < 1.1$ are considered relaxed whereas objects with $A_{180} > 1.1$ are disturbed \citep{Zhang2010,Okabe2010}.

Combining these 2 measurements along with by-eye inspection we are able to accurately define the physical state of the galaxy clusters in our sample. 

\subsection{Strong Lensing Models and Core Lensing Mass Measurement}
\label{sec:clm}

The strong lensing models for these galaxy clusters were derived in a series of papers by Sharon and colleagues \cite{Sharon2020,Sharon2022a,Sharon2022b} using the publicly available Lenstool software \cite{Jullo2007}. We use 100 lens models drawn from the \texttt{Lenstool} MCMC with the derived lensing parameters to sample the posterior probability distribution. Though strong lensing allows us to study the distribution at small scales, the obvious core cluster that dominates the distribution at large scales is the component we are interested in since we would expect it to align with the BCG, ICL, and ICM components. The parameters from the most likely image plane model are taken to be the true values of the core lensing mass.

\begin{figure}[ht!]
	\includegraphics[width=\columnwidth]{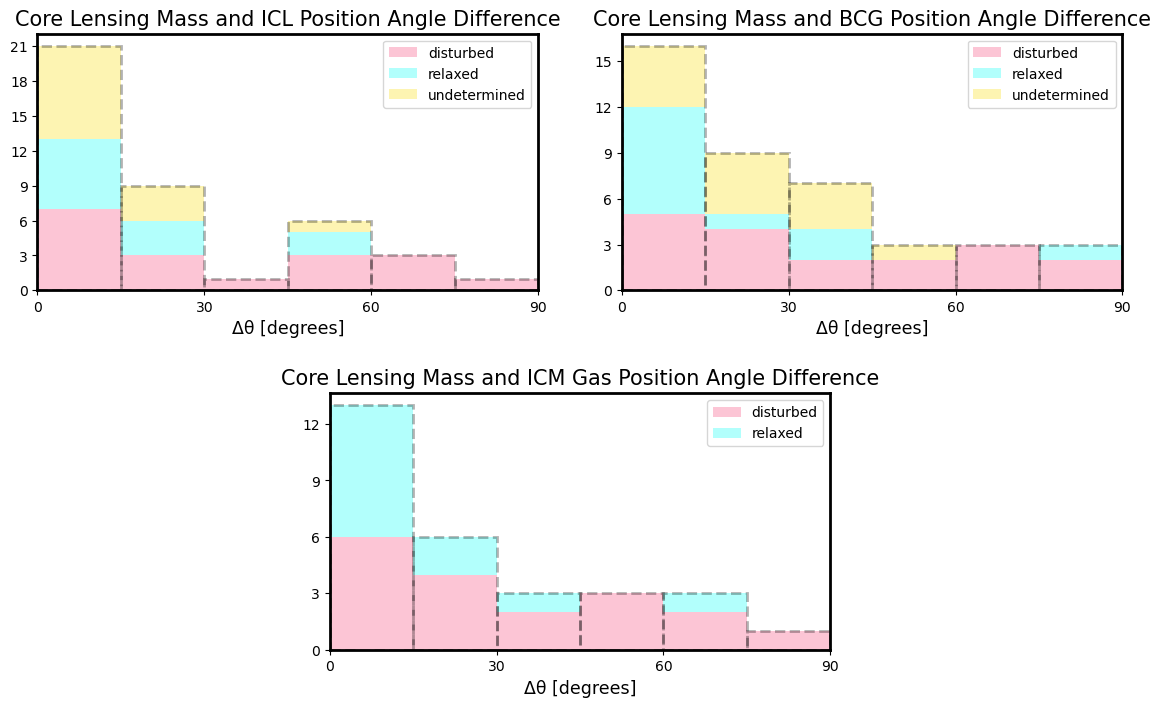}
    \caption{The difference in position angle of the major axis between the core lensing mass and the three other distributions (from top to bottom: ICL, BCG, and ICM Gas). The histogram is color-coded to indicate the dynamical state of the cluster with the undetermined label indicating objects without \emph{Chandra} data.}
    \label{fig:pa_mass_sum_23.png}
\end{figure}

\section{Results}
\label{sec:results}

\subsection{Position Angles}
\label{sec:pa}

Figure~\ref{fig:pa_mass_sum_23.png} shows the various cluster components’ position angle differences with respect to the core lensing mass. Equation~\ref{eq:pa} defines the difference in position angle.

\begin{align}
\label{eq:pa}
\begin{split}
 \Delta\mathrm{PA}=|\mathrm{PA}_1-\mathrm{PA}_2| \;\;\;\;\;\;\;\;\;
 (|\mathrm{PA}_1-\mathrm{PA}_2|\le90^\circ)
\\
 \Delta\mathrm{PA}=180-|\mathrm{PA}_1-\mathrm{PA}_2| \;\;\;\;\;\;\;\;\;
 (|\mathrm{PA}_1-\mathrm{PA}_2|>90^\circ)
\\
0^\circ \le \Delta\mathrm{PA}\le90^\circ \;\;\;\;\;\;\;\;\;\;\;\;\;\;\;\;\;\;\;\;\;\;\;\;\;\;\;\;\;\;\;\;\;\;\;\;\;\;\;\;\;
\end{split}
\end{align}

For disturbed and relaxed objects alike, we mostly measure small position angle differences which implies that cluster orientation is maintained from a few tens of kpc up to $\sim$1Mpc. This consistency over large spatial scales has been previously observed \citep[e.g.][]{Donahue2016}.

Still, we find multiple instances of large position angle differences ($\Delta$PA$>30^{\circ}$), more so for disturbed clusters than relaxed ones.
All components are sensitive to dynamical disruptions. However, the relatively small number of large position angle differences between the core lensing mass and ICL suggests that the ICL may be the best proxy for the DM halo distribution, consistent with previous studies \citep[e.g.][]{Montes2018}.

\begin{figure}[ht!]
\begin{center}
	\includegraphics[width=.8\columnwidth]{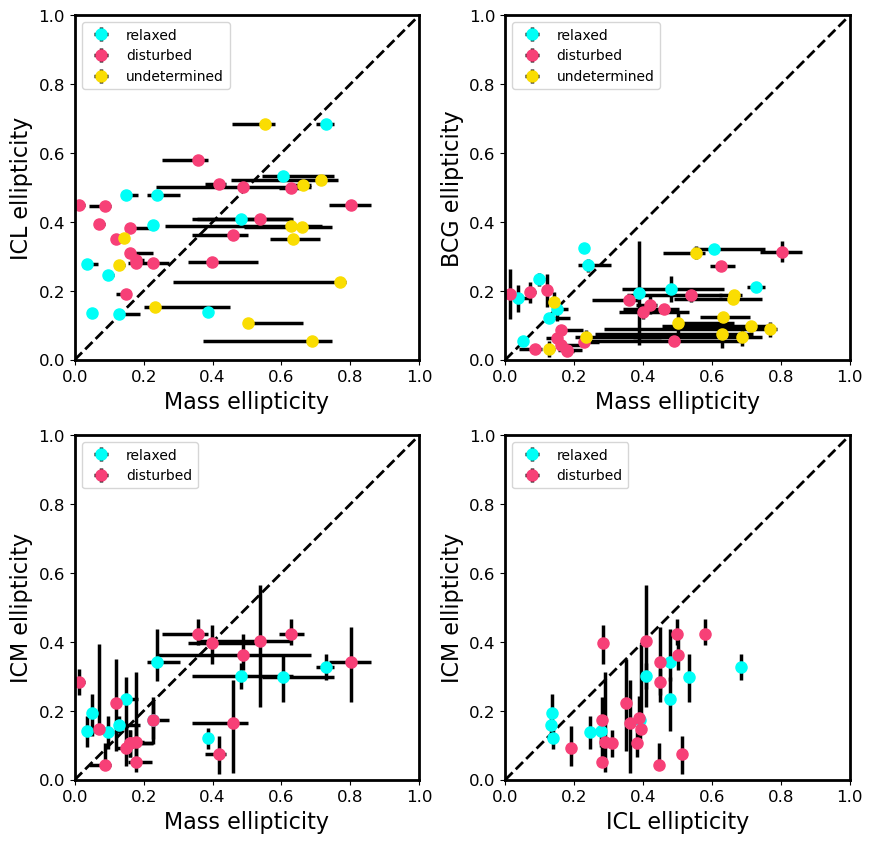}
    \caption{The ellipticity comparisons of various distributions (top-left: core lensing mass and ICL, top-right: core lensing mass and BCG, bottom-left: core lensing mass and ICM, bottom-right: ICL and ICM). The histogram is color-coded to indicate the dynamical state of the cluster with the undetermined label indicating objects without \emph{Chandra} data.}
    \label{fig:eps_mass_sum23}
\end{center}
\end{figure}

\subsection{Ellipticities}

\label{sec:eps}

Figure~\ref{fig:eps_mass_sum23} compares the ellipticity measurements for the various cluster components. We define ellipticity as the flattening parameter described by $e=1-{b}/{a}$, where a and b are the semi-major and semi-minor axes.

Overall, there is no clear ellipticity trends based on degree of relaxation. From the top-left panel, we see that the DM and the ICL have roughly similar ellipticities with a large scatter. As seen in the bottom 2 panels, the ICM is rounder than the DM or ICL components due to hydrodynamical effects occurring in the ICM \citep{Markevitch2007}. In the top-right panel, we see the BCG is more circular due to dynamical friction caused by high stellar density \citep*{Arena2006}.

\begin{figure*}[ht!]
	\includegraphics[width=.95\columnwidth]{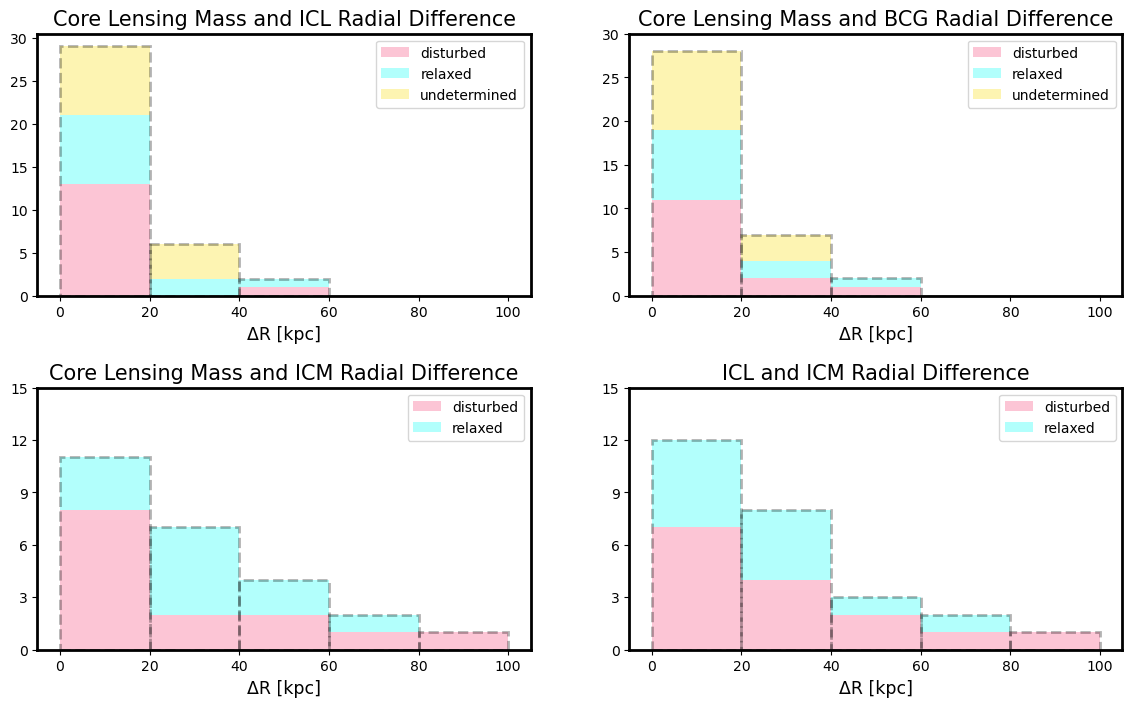}
    \caption{The centroid comparisons of various distributions (top-left: core lensing mass and ICL, top-right: core lensing mass and BCG, bottom-left: core lensing mass and ICM, bottom-right: ICL and ICM). We exclude doubly cored systems with extremely large deviations for the purpose of visualization. The histogram is color-coded to indicate the dynamical state of the cluster with the undetermined label indicating objects without \emph{Chandra} data.}
    \label{fig:cent_mass_sum23}
\end{figure*}

\subsection{Centroids}
\label{sec:cent}

From figure~\ref{fig:cent_mass_sum23}, we see the centroid difference comparisons of the various components of our galaxy clusters. We use the difference in projected radius as defined by $\Delta\mathrm{R}=|r_1-r_2|$ in units of kpc to quantify the difference in centroid.

As expected, we measure small deviations in centroid for the ICL and BCG when compared to the core lensing mass which illustrates the typical alignment of the stellar components with respect to the center of the DM potential. The ICL is very slightly more aligned with the core lensing mass centroid than the BCG since the BCG experiences a greater residual wobble. 

Even after excluding obvious major merger systems with two distinct/significantly radially displaced cores, we still see that the ICM Gas is displaced much more than the ICL or BCG when compared to the core lensing mass. This may happen since major merger activity in the cluster’s past can cause ICM “sloshing” \cite{Markevitch2001,Churazov2003,Johnson2012,Harvey2017}. The sloshing behavior can persist even when the stellar components have relaxed back the to DM centroid since it is a consequence of hydrodynamical physics that affects the ICM only \cite{Markevitch2007}.

Surprisingly, we do not see a tendency for the BCG or ICL to have much larger radial displacements for disturbed clusters. This could be due to the stellar components having relaxed back to the DM centroid but experiencing displacements due to small scale astrophysics as opposed to merger activity. For the ICM, there is a slight preference to larger displacements for disturbed clusters due to the ICM's sensitivity to mergers.

\bibliography{main.bib}

\end{document}